\documentclass{article}

\usepackage[utf8]{inputenc}
\usepackage[T1]{fontenc}
\usepackage{graphicx}
\usepackage{amsmath,amssymb}
\usepackage{amsthm}
\usepackage{subcaption}
\usepackage{hyperref}
\usepackage{listings}
\usepackage{mathtools}
\usepackage{xcolor}
\usepackage{array}
\newcolumntype{L}[1]{>{\raggedright\let\newline\\\arraybackslash\hspace{0pt}}m{#1}}
\newcolumntype{C}[1]{>{\centering\let\newline\\\arraybackslash\hspace{0pt}}m{#1}}
\newcolumntype{R}[1]{>{\raggedleft\let\newline\\\arraybackslash\hspace{0pt}}m{#1}}

\newcommand{\old}[1]{\textcolor{red}{\ifmmode\text{\sout{\ensuremath{#1}}}\else\sout{#1}\fi}}

\newcommand{\worldtype}{\ensuremath{\mu}} 
\newcommand{\accrel}{\ensuremath{r}} 
\newcommand{\truthtypehol}{\ensuremath{o}} 
\newcommand{\eiw}{\ensuremath{\text{eiw}}} 
\newcommand{\tte}{\ensuremath{\sigma}} 

\begin{document}

\title{Solving Quantified Modal Logic Problems\\by Translation to Classical Logics}
\author{Alexander Steen \and Geoff Sutcliffe \and Christoph Benzm{\"u}ller}

\maketitle

\begin{abstract}
This article describes an evaluation of Automated Theorem Proving (ATP) systems on problems
taken from the QMLTP library of first-order modal logic problems.
Principally, the problems are translated to both typed first-order and higher-order logic
in the TPTP language using an embedding approach, and solved using first-order resp.\
higher-order logic ATP systems and model finders.
Additionally, the results from native modal logic ATP systems are considered, and compared
with the results from the embedding approach.
The findings are that the embedding process is reliable and successful when state-of-the-art ATP
systems are used as backend reasoners,
The first-order and higher-order embeddings perform similarly,
native modal logic ATP systems have comparable performance to classical systems
using the embedding for proving theorems, 
native modal logic ATP systems are outperformed by the embedding approach for disproving
conjectures, and the embedding approach can cope with a wider range of modal logics than the native
modal systems considered.

\end{abstract}



\section{Introduction}
\label{Introduction}

Automated Theorem Proving (ATP) systems try to prove formally and fully automatically that a
conjecture is entailed by a given set of premises.
For first-order quantified modal logics \cite{FM98} there exist a few ATP systems, including
GQML \cite{TCC02}, MleanTAP, MleanSeP, MleanCoP \cite{Ott14}, and the more recent nanoCoP-M 2.0
\cite{Ott21}.
Unfortunately, none of them support all 15 modal logics in the modal cube \cite{Gar18}.
For example, MleanTAP, MleanCoP, and nanoCoP-M 2.0 support only the \textbf{D}, \textbf{M} 
(aka \textbf{T}), \textbf{S4}, and \textbf{S5} logics; GQML supports only \textbf{K}, \textbf{D}, 
\textbf{K4}, \textbf{M}, and \textbf{S4}; and MleanSeP supports only \textbf{K}, \textbf{K4}, 
\textbf{D}, \textbf{D4}, \textbf{S4}, and \textbf{M}.
As a consequence, even with the most versatile (in terms of the number of different modal logics 
supported) ATP systems from this list only 40\% of all modal logics from the modal cube are covered.
This effect is multiplied when taking into account further modal logic properties such as
different domain semantics \cite{FM98}, rigidity of terms \cite{BBW06}, and multi-modal
combinations \cite{BBW06} of these properties.
For example, non-rigid terms are covered by only GQML, and none of the listed systems support
decreasing domains (also referred to as anti-monotonic frames \cite{FM98}).
Further notions like global and local assumptions \cite{FM98,BdV01} are unsupported,
but are important, e.g., for applications in theoretical philosophy \cite{BW16}.

An alternative to developing native modal logic ATP systems is to translate modal logic problems
to classical first-order logic \cite{Ohl91,BG07} or higher-order logic \cite{BOR12,BR13,BP13}, and then 
solve the first- resp.\ higher-order logic problems. 
Translations to higher-order logic were primarily motivated by the automation of modal logics with 
propositional, second-order or generally higher-order quantification; they turned out to provide 
convincing automation for standard first-order modal logics with acceptable performance and, in 
particular, unchallenged flexibility~\cite{GSB17,Ben19,SS+23}.
In particular, this indirect approach provides a rich ecosystem for adjusting the different properties of the modal
logic under consideration.
In this article three state-of-the-art ATP systems and one model finder are evaluated on
translations of first-order modal logic problems taken from the QMLTP library \cite{RO12}.
Earlier results, also for other classical ATP systems, are presented in \cite{BOR12,BR13,GSB17,SS+23}.
Over the last decade there has been substantial progress in the development of reasoning systems
for classical first-order and higher-order logic, including those used in this work:
E 3.0.03~\cite{SCV19,VB+21},
Leo-III 1.7.8~\cite{SB18},
Nitpick 2016~\cite{BN10-ITP},
and 
Vampire 4.8~\cite{BR20-SUP}.

Extending earlier work~\cite{SS+23}, the goals of this work are:
\begin{enumerate}
\item Test the process of translating first-order modal logic problems into classical first-
      and higher-order logic using a shallow embedding.
\item Evaluate multiple first- and higher-order ATP systems' abilities to solve the embedded
      modal logic problems.
\item Compare the performance of native modal logic ATP systems with the embedding
      approach.
\end{enumerate}
The first goal provides feedback regarding the process, ensuring that the classical first- and
higher-order logic problems faithfully reflect the original non-classical modal logic problems.
In particular, the soundness and practicality of this translation process is tested by a mutual 
comparison of prover results; discrepancies in prover results would indicate potential errors.
The second goal provides information about which first- and higher-order ATP systems can be most
effectively used as a backend in a tool chain that solves modal logic problems by translation
to first- resp.\ higher-order logic.
The third goal provides an interesting evaluation of the efficacy of the embedding approach
in comparison to reasoning directly in modal logic.

The remainder of this article is structured as follows:
Section~\ref{Prelim} briefly introduces modal logic, the QMLTP library, and higher-order logic.
Section~\ref{TPTPLanguages} introduces the TPTP languages that have been used in this work
for encoding the modal logic problems and their corresponding translations to classical logic.
Section~\ref{Embedding} describes how non-classical logic problems can be translated to
classical first- and higher-order logic problems, with a particular emphasis on modal logic
as used in the QMLTP.
Section~\ref{Evaluation} describes the experimental setup, and the results of the evaluation.
A performance comparison with native modal logic ATP systems is provided.
Section~\ref{Conclusion} concludes and sketches further work.

\section{Preliminaries}
\label{Prelim}

\subsection{Modal Logic}\label{ModalLogic}

First-order modal logic (FOML) is a family of logic formalisms extending classical first-order
logic (FOL, here without equality) with the unary modal operators $\Box$ and $\Diamond$.
FOML terms are defined as for FOL.
The modal logic \textbf{K} is the smallest logic that includes all FOL tautologies, all instances
of the axiom scheme K ($\Box(\varphi \rightarrow \psi) \rightarrow
(\Box\varphi \rightarrow \Box\psi)$), and is closed with respect to modus ponens (from
$\varphi \rightarrow \psi$ and $\varphi$ infer $\psi$) and necessitation (from
$\varphi$ infer $\Box \varphi$).
In normal modal logics the modal operators are dual notions, i.e., $\Box \varphi$ is equivalent
to $\neg \Diamond \neg \varphi$.
Further axiom schemes are assumed for stronger modal logics \cite{FM98}.

\textbf{K} can alternatively be characterized by Kripke structures~\cite{Kri63}.
For a FOML language over signature $\Sigma$, a first-order Kripke frame is a tuple
$M = (W, R, \mathcal{D}, \mathcal{I})$, where $W$ is a non-empty set (the possible worlds),
$R \subseteq W \times W$ is a binary relation on $W$ (the accessibility relation),
$\mathcal{D} = \{D_w\}_{w \in W}$ is a family of non-empty sets $D_w$ (the domain of world $w$)
and $\mathcal{I} = \{ I_w \}_{w \in W}$ is a family of interpretation functions $I_w$, one for
each world $w \in W$, that map the symbols of $\Sigma$ in world $w \in W$ to adequate denotations
over $D_w$.
In Kripke semantics the truth of a formula $\phi$ with respect to $M$ and a world $w \in W$,
written $M,w \models \phi$, is defined as usual~\cite{FM98}.
For stronger modal logic systems, restrictions may be imposed on the relation $R$, e.g.,
modal logic \textbf{D} is characterized by the class of Kripke frames where $R$ is serial.
Similar correspondence results exist for the other common modal logics~\cite{Gar18}.
Multi-modal logics generalize the language with multiple (indexed) modalities, where
the accessibility relation $R$ is replaced with a set of relations
$\mathcal{R} = \{ R_i \}_{i \in I}$, one for each indexed modality $\Box_i$, $i \in I$,
where $I$ is some index set.
Each indexed modality can be assumed to satisfy stronger properties by restricting its
accessibility relation $R_i$.

Following Fitting and Mendelsohn~\cite{FM98}, local consequence of a formula $\varphi$ in FOML
is defined with respect to a set of global assumptions $G$ and a set of local assumptions $L$,
written $G \models L \rightarrow \varphi$.
If $G$ is empty, this reduces to the common notion of local consequence.
See \cite{FM98} for the precise definition.

Further variations of the Kripke semantics are possible; three independent refinements are
well-discussed in the literature.
Firstly, quantification semantics may be specialized by domain restrictions \cite{FM98}:
In \emph{constant domains} semantics (also called possibilist quantification), all domains are assumed
to coincide, i.e., $D_w = D_v$ for all worlds $w,v \in W$.
In \emph{cumulative domains} semantics, the domains are restricted such that $D_w \subseteq D_v$
whenever $(w,v) \in R$, for all $w,v \in W$.
In \emph{decreasing domains} semantics, it holds that $D_v \subseteq D_w$ whenever $(w,v) \in R$,
for all $w,v \in W$.
In \emph{varying domains} semantics (also called actualist quantification) no restriction is imposed
on the domains.
These semantic restrictions can be equivalently characterized in a proof-theoretic way using the
(converse) Barcan formulae \cite{Bar46,FM98}.
Secondly, constancy restrictions of the interpretation of constant and function symbols across different possible
worlds may be applied.
The interpretation is \emph{rigid} if $I_w(c) = I_v(c)$
and $I_w(f) = I_v(f)$ for each constant symbol $c \in \Sigma$ and function symbol $f \in \Sigma$,
and all worlds $w,v \in W$.
If this is not the case the interpretation is \emph{flexible}. This distinction is also sometimes
referred to as rigid vs.\ flexible (or: world-dependent) \emph{designation}. Details are discussed in \cite{FM98}.
Thirdly, the domain of admissible interpretations of terms may be constrained. So-called \emph{local terms}~\cite{RO12} require
that, for every world $w$, every ground term $t$ denotes at $w$ an object that exists at $w$, in particular
it holds that $\mathcal{I}_w(c) \in \mathcal{D}_w$ and $\mathcal{I}_w(f)(d_1, \ldots, d_n) \in \mathcal{D}_w$
for every world $w$, every constant symbol $c \in \Sigma$, every function symbol $f \in \Sigma$ of arity $n$, and objects
$d_1, \ldots, d_n \in \mathcal{D}_w$. If such a restriction is not imposed, terms are sometimes referred to as \emph{global}.
Local terms should not be confused with local assumptions (local consequence): The former speaks about restrictions
on the interpretation of terms, the latter about the the consequence relation.

\subsection{The QMLTP library}

The Quantified Modal Logics Theorem Proving (QMLTP) library\footnote{%
\href{http://www.iltp.de/qmltp}{\tt http://www.iltp.de/qmltp}}
provides a platform for testing and benchmarking ATP systems for FOML.
It is modelled after the TPTP library for classical logic \cite{Sut17}.
It includes a problem collection of FOML problems, information about ATP systems for FOML, and
performance results from the systems on the problems in the library (last updated around 2012).
The problem collection is divided into 11 domains in five groups:
\begin{itemize}
\item The {\tt APM} domain of ``mixed applications''.
      There are 10 problems.
\item The {\tt G??} domains are G{\"o}del encodings of TPTP problems viewed as intuitionistic
      problems.
      There are 245 problems.
\item The {\tt MML} multi-modal logic problems.
      There are 20 problems, each with a particular logic specification.
\item The {\tt NLP} and {\tt SET} domains of classical logic problems, treated as modal logic
      problems.
      They have no modal connectives, but are are included in the evaluation because they are 
      part of the QMLTP.
      There are 5 {\tt NLP} and 75 {\tt SET} problems.
\item The {\tt SYM} domain of syntactic modal logic problems.
      There are 245 problems.
\end{itemize}
Overall there are 600 problems, of which 580 are mono-modal and 20 are multi-modal.
The problems assume rigid constants, local terms, and local consequence (i.e., all 
assumptions in the problem files are local).

Each problem is documented with the expected result for all combinations of constant, 
cumulative, and varying domains, in modal systems {\bf K}, {\bf D}, {\bf M}, {\bf S4}, and 
{\bf S5}.
The problems do not have expected results for decreasing domains, assumedly because of the absence
of native modal logic ATP systems that support decreasing domains.
The FOML framework from Section~\ref{ModalLogic} subsumes these particular choices.
In general, each problem might have a different expected result for each combination of logic 
properties (i.e., being a theorem or not).
There are problems that are unprovable for any combination of modal logic properties, problems 
that are provable in some but not all combinations, and problems that are provable in all 
combinations of properties considered in the QMLTP.

The QMLTP library also defines a lightweight syntax extension of the TPTP's FOF language, 
denoted {\tt qmf}, for expressing FOML.
The {\tt qmf} syntax is not introduced here, and instead the TPTP language for modal logic, 
introduced in Section~\ref{TPTPLanguages}, is used.

\subsection{Higher-Order Logic}

There are many quite different frameworks that fall under the general label ``higher-order''.
The notion reaches back to Frege's original predicate calculus \cite{Fre93-03}.
Church 
introduced simple type theory \cite{Chu40}, a higher-order
framework built on his simply typed $\lambda$-calculus, employing types to reduce expressivity
and to remedy paradoxes and inconsistencies.
Variants and extensions of Church's simple type theory have been the logic of choice for
interactive proof assistants such as HOL4 \cite{GM93}, HOL Light \cite{Har96-FMCAD},
PVS \cite{OR+96}, Isabelle/HOL \cite{NPW02}, and OMEGA \cite{SBA06}.
A variation of simple type theory is extensional type theory \cite{Hen50,BM14}.
It is a common basis for higher-order ATP systems, including those used in this work.
For the remainder of this article ``higher-order logic'' (HOL) is therefore synonymous with
extensional type theory, and is the intended logic of the TPTP THF language \cite{SB10} used
in this work (see Section~\ref{TPTPLanguages}).
The semantics is the general semantics (or Henkin semantics), due to Henkin and Andrews
\cite{Hen50,And72,BBK04}.
See \cite{BM14} for a full introduction to higher-order logic syntax and semantics.

\section{The TPTP Languages}
\label{TPTPLanguages}

The TPTP languages \cite{Sut23-IGPL} are human-readable, machine-parsable, flexible and 
extensible languages, suitable for writing both ATP problems and ATP solutions\footnote{%
The development of TPTP World standards for writing ATP solutions beyond common
derivations and models is still necessary; see, e.g., \cite{OS10}}.
In this section the general structure of the TPTP languages is reviewed, and the
two specific TPTP languages used for non-classical logics are presented.
The full syntax of the TPTP languages is available in extended BNF form.\footnote{%
\href{http://www.tptp.org/TPTP/SyntaxBNF.html}{\tt http://www.tptp.org/TPTP/SyntaxBNF.html}}

The top-level building blocks of the TPTP languages are {\em annotated formulae},
in the form:\\
\hspace*{1cm}{\em language}{\tt (}{\em name}{\tt ,} {\em role}{\tt ,} {\em formula}{\tt ,}
{\em source}{\tt ,} {\em useful\_info}{\tt ).} \\
The {\em language}s supported are clause normal form ({\tt cnf}), first-order form ({\tt fof}),
typed first-order form ({\tt tff}), and typed higher-order form ({\tt thf}).
{\tt tff} and {\tt thf} are each used for multiple languages in the TPTP language hierarchy, 
described below.
The {\em name} assigns a (unique) identifier to each formula, for referring to it.
The {\em role}, e.g., {\tt axiom}, {\tt lemma}, {\tt conjecture}, defines the use of the formula
in an ATP system.
In the {\em formula}, terms and atoms follow Prolog conventions.
The TPTP language also supports interpreted symbols, including: 
``the type of types'' {\tt \$tType};
types for individuals {\tt \$i} ($\iota$) and booleans {\tt \$o} ($o$);
types for numbers {\tt \$int} (integers), {\tt \$rat} (rationals), and {\tt \$real} (reals);
numeric constants such as 27, 43/92, -99.66;
arithmetic predicates and functions such as {\tt \$greater} and {\tt \$sum};
the truth constants {\tt \$true} and {\tt \$false}.
The basic logical connectives are {\tt \verb|^|}, {\tt !}, {\tt ?}, {\tt {@}}, 
{\tt {\raisebox{0.4ex}{\texttildelow}}}, {\tt |}, {\tt \&}, {\tt =>}, {\tt <=}, {\tt <=>}, and 
{\tt <{\raisebox{0.4ex}{\texttildelow}}>},
for $\lambda$, $\forall$, $\exists$, higher-order application, $\neg$, $\vee$, $\wedge$, 
$\Rightarrow$, $\Leftarrow$, $\Leftrightarrow$, and $\oplus$ respectively.
Equality and inequality are expressed as the infix operators {\tt =} and {\tt !=}.
The {\em source} and {\em useful\_info} are optional extra-logical information about the origin 
and useful details about the formula.
See \cite{Sut17} or the TPTP web site \href{https://www.tptp.org}{\tt https://www.tptp.org} for 
all the details.
An example of an annotated first-order formula defining the set-theoretic union operation,
supplied from a file named {\tt SET006+1.ax}, is~\ldots
\[
\begin{minipage}{\textwidth}
\begin{verbatim}
    fof(union,axiom,
        ( ! [X,A,B] :
            ( member(X,union(A,B))
          <=> ( member(X,A) | member(X,B) ) ),
        file('SET006+0.ax',union),
        [description('Definition of union'), relevance(0.9)]).
\end{verbatim}
\end{minipage}
\]

The TPTP has a hierarchy of languages that ends at the non-classical languages used in this work.
The languages are:
\begin{itemize}
\item Clause normal form (CNF), which is the ``assembly language'' of many modern ATP systems.
\item First-order form (FOF), which hardly needs introduction.
\item Typed first-order form (TFF), which adds types and type signatures, with monomorphic (TFF)
      and polymorphic (TF1) variants.
\item Typed extended first-order form (TXF), which adds Boolean terms, Boolean variables as
      formulae, tuples, conditional expressions, and let expressions.
      TXF has monomorphic (TX0) and polymorphic (TX1) variants.
\item Typed higher-order form (THF), which adds higher-order notions including curried type
      declarations, lambda terms, partial application, and connectives as terms.
      THF has monomorphic (THF) and polymorphic (TH1) variants.
      THF is the TPTP language used for HOL.
\item Non-classical forms (NXF and NHF), which add {\em non-classical connectives} and
      {\em logic specifications}.
      The non-classical typed extended first-order form (NXF) builds on TXF, and the
      non-classical typed higher-order form (NHF) builds on THF.
      NXF and NHF are the TPTP languages used for FOML.
\end{itemize}

\subsection{Non-Classical Connectives}
\label{Connectives}

The non-classical connectives of NXF and NHF have the form
{\tt \verb|{|\$}{\em connective\_name}{\tt \verb|}|}.
A connective may optionally be parameterized to reflect more complex non-classical connectives,
e.g., in multi-modal logics where the modal operators are indexed, or in epistemic logics
\cite{vDH15} where the common knowledge operator can specify the agents under consideration.
The form is
{\tt \verb|{|\$}{\em connective\_name}{\tt (}{\em param$_1$}{\tt ,}{\em \ldots}{\tt ,}{\em param$_n$}{\tt )}{\tt \verb|}|}.
If the connective is indexed the index is given as the first parameter prefixed with a {\tt \#}.
All other parameters are key-value assignments.
In NXF the non-classical connectives are applied in a mixed 
``higher-order applied''/``first-order functional'' style, with the connectives applied to a
{\tt ()}ed list of arguments.\footnote{%
This slightly unusual form was chosen to reflect the first-order functional style, but by making
the application explicit the formulae can be parsed in Prolog -- a long standing principle of the
TPTP languages \cite{SZS04}.}
In NHF the non-classical connectives are applied in usual higher-order style, with curried 
function applications using the application operator \verb|@|.
For FOML the connectives are {\tt \{\$box\}} and {\tt \{\$dia\}} for $\Box$ and $\Diamond$.
In the context of multi-modal FOML they are {\tt \{\$box(\#i)\}} and {\tt \{\$dia(\#i)\}} for
$\Box_i$ and $\Diamond_i$, respectively, where {\tt \#i} is a representation of an index $i$.
Figure~\ref{NCLLongFormExample} illustrates the use of connectives from some
(not further specified), modal, alethic modal, and epistemic logics.

\begin{figure}[tb]
\caption{Non-classical connective examples} \label{NCLLongFormExample}
\centering
\scriptsize
\begin{subfigure}[t]{.48\textwidth}
\begin{verbatim}
tff(pigs_fly_decl,type, pigs_fly: $o ).  

tff(flying_pigs_impossible,axiom,
    ~ {$possible} @ (pigs_fly) ). 
          
tff(alice_knows_pigs_dont_fly,axiom,
    {$knows(#alice)} @ (~ pigs_fly) ).

tff(something_is_necessary,axiom,
    ? [P: $o] : {$necessary} @ (P) ).


tff(a_decl,type, a: $i ).
tff(f_decl,type, f: $i > $o ).

tff(reasonable,conjecture,
    ( ! [X: $i] : ( {$box} @ (f(X)) )
   => ( {$box} @ (! [X: $i] : f(X)) ) ) ).

tff(silly,conjecture,
    ( {$box(#a)}
    @ ( {$dia(#a)} @ (! [X: $i] : f(X)) ) ) ).
\end{verbatim}
\end{subfigure}
\begin{subfigure}[t]{.48\textwidth}
\begin{verbatim}
thf(positive_decl,type,
    positive: ($i > $o) > $o ).

thf(self_identity_is_positive,axiom,
    {$necessary} @
      ( positive @ ^ [X:$i] : (X = X) ) ).

thf(alice_and_bob_know,axiom,
    {$common($agents:=[alice,bob])} @
      ( positive @ ^ [X:$i] : (X = X) ) ).

thf(everything_is_possibly_positive,axiom,
    ! [P: $i > $o] :
      ( {$possible} @ (positive @ P) ) ).
\end{verbatim}
\end{subfigure}
\end{figure}

\subsection{Logic Specifications}
\label{Specifications}

In the world of non-classical logics the intended logic cannot be inferred from the language used 
for the formulae -- the same syntactical language that is used for representing formulae
can host different logics with different proof-theoretic inferences on the formulae.
For example, when reasoning about metaphysical necessity modal logic \textbf{S5} is usually
used, but when reasoning about deontic necessities a more suitable choice might be modal logic
\textbf{D}. Nevertheless, modal logics \textbf{S5} and \textbf{D} share the same logical language.
It is therefore necessary to provide {\mbox{(meta-)}information} that specifies the logic to
be used.

A new kind of TPTP annotated formula is used to specify the logic and its properties, 
The annotated formulae has the role \texttt{logic}, and has a ``logic specification'' as its
formula.
A logic specification consists of a defined logic (family) name identified with a list of
properties, e.g., in NXF~\ldots\\
\hspace*{1cm}{\tt tff(}{\em name}{\tt,logic,}{\em logic\_name} {\tt ==} {\em properties}{\tt ).} \\
where {\em properties} is a {\tt \verb|[]|} bracketed list of key-value identities~\ldots\\
\hspace*{1cm}{\em property\_name} {\tt ==} {\em property\_value} \\
where each {\em property\_name} is a TPTP defined symbol or system symbol,
and each {\em property\_value} is either a term (often a defined constant) or a
{\tt \verb|[]|} bracketed list that might start with a term (often a defined constant), and 
otherwise contains key-value identities.
If the first element of a {\em property\_value} list is a term then that is the default value
for all cases that are not specified by the following key-value identities.

In this work logic specifications for FOML are used.
The TPTP reserves the logic identifier {\tt \$modal} for modal logics.
The {\tt \$modal} family follows the generalized notion of consequence for modal logics by
Fitting and Mendelsohn \cite{FM98}, as briefly introduced in Section~\ref{ModalLogic},
which allows for both local and global premises in problems \cite[Ch~1.5]{BdV01}.
Annotated formulae with the {\tt axiom} role are global premises (true in all worlds),
and those with the {\tt hypothesis} role are local premises (true in the current world).
These default role readings can be overridden using {\em subroles}, e.g., a formula with the role
{\tt axiom-local} is local, and a formula with the role {\tt hypothesis-global} is global.
The following properties can be specified for {\tt \$modal} logics
(as discussed in Section~\ref{ModalLogic}):
\begin{itemize}
\item The \verb|$domains| property specifies restrictions on the domains across the accessibility 
      relation.
      The possible values are {\tt \$constant}, {\tt \$varying}, {\tt \$cumulative}, and
      {\tt \$decreasing}.
\item The \verb|$designation| property specifies whether {\em symbols} are interpreted as 
      {\tt \$rigid}, i.e., interpreted as the same domain element in every world, or as 
      {\tt \$flexible}, i.e., possibly interpreted as different domain elements in different
      worlds.
\item The \verb|$terms| property specifies whether local term restrictions apply
      (value {\tt \$local}) or not (value {\tt \$global}).
\item The \verb|$modalities| property specifies the modality of {\em connective}(s), either
      all connectives, or by index.
      Possible values are defined for well-known modal logic systems,
      e.g., {\tt \$modal\_system\_K}, and lists of individual modal axiom schemes, e.g.,
      {\tt \$modal\_axiom\_5}.
      They refer to the corresponding systems and axiom schemes of the modal logic cube
      \cite{Gar18}.
\end{itemize}
In order to reduce misunderstandings there are no default values. 
It is an error if a problem does not specify all relevant properties (but irrelevant ones may 
be omitted, e.g.,\ \verb|$domains| may be omitted if the problem does not contain any 
quantification).

A simple example from modal logic \textbf{S5} with constant domains except for variables of 
{\tt some\_user\_type}, local terms, and with rigid constants, is~\ldots
\[
\begin{minipage}{\textwidth}
\begin{verbatim}
   tff(simple_spec,logic,
       $modal == [
         $domains == [ $constant, some_user_type == $varying ],
         $designation == $rigid,
         $terms == $local,
         $modalities == $modal_system_S5 ] ).
\end{verbatim}
\end{minipage}
\]
Multi-modal logics are specified by enumerating the modalities of the connectives.
Here is an example of a logic specification with constant domains, rigid symbols, and local terms.
Modalities are generally {\bf K}, but $\Box_a$ satisfies modal axiom K and B, and $\Box_b$ is a 
\textbf{S4} modality.
\[
\begin{minipage}{\textwidth}
\begin{verbatim}
   thf(multi_spec,logic,
       $modal == [
         $domains == $constant,
         $designation == $rigid,
         $terms == $local 
         $modalities == [ $modal_system_K,
           {$box(#a)} == [ $modal_axiom_K, $modal_axiom_B ],
           {$box(#b)} == $modal_system_S4 ] ] ).
\end{verbatim}
\end{minipage}
\]

\section{Shallow Embeddings into Classical Logic}
\label{Embedding}

Shallow embeddings of a source logic into a target logic are translations that encode the 
semantics of formulae of the source logic as terms or formulae of the target logic.
This is in contrast to so-called deep embeddings that encode the source logic formulae as
uninterpreted data (usually an inductively defined datatype), and define meta-theoretical
notions such as interpretation and satisfiability as functions and predicates.
An in-depth discussion of pros and cons of both approaches is presented in \cite{GW14}.
The shallow embedding of modal logics into classical higher-order logic 
used in this work is based on earlier work~\cite{BOR12,BP13,BR13,GSB17}. 
The shallow embedding of modal logics into classical (typed) first-order logic is commonly referred to
as \emph{standard translation}~\cite{Ohl91,BG07} and widely known.
The main ideas of the embedding into higher-order logic are informally sketched in the following.
The embedding into first-order logic is conceptually very similar and based on the standard translation
of~\cite{BG07}, which is not explained in detail here. The first-order embedding was
suitably augmented to cover all the different modal logic parameters, as explained below.

The notion of Kripke semantics~\cite{BdV01,BBW06} is simulated in HOL as follows:
A new type $\worldtype$ is introduced as the type of possible worlds, and a binary relation (predicate) symbol
$\accrel^i_{\worldtype \rightarrow \worldtype \rightarrow o}$, where $o$ is the classical type of
truth values in HOL, models the (indexed) accessibility relation
$R^i$ between worlds.
Since the truth of a formula $\varphi$ in FOML is established through assessing the truth of
$\varphi$ with respect to all worlds, formulae of FOML are identified with HOL predicates of
type $\worldtype \rightarrow \truthtypehol$, abbreviated $\tte$ in the following.
Connectives are defined accordingly, evaluating a composite formula with respect to a given world.
For example, the box operator, negation, disjunction, and constant-domain universal quantification
over objects of type $\tau$ are defined as follows
(subscripts denote types: a type $\tau \to \nu$ is the type of total functions from objects
of type $\tau$ to objects of type $\nu$;
expressions of form $\lambda X_\tau. T_\nu$ are anonymous functions mapping their parameters
$X_\tau$ to some object of type $\nu$ as given by term $T$):
\begin{equation*}\begin{array}{ll}
{\Box}_{{\tte \rightarrow {\tte}}}^{{i}} &:= {\lambda S_\tte. \lambda W_\worldtype. \forall V_\worldtype. \; \neg (r^i \; W \; V) \lor S \; V} \\
{\neg}_{{{\tte} \rightarrow {\tte}}} &:= {\lambda S_\tte. \lambda W_\worldtype. \; \neg (S \; W)} \\
{\lor}_{{{\tte} \rightarrow \tte \rightarrow {\tte}}} &:= {\lambda S_\tte. \lambda T_\tte. \lambda W_\worldtype. \; (S \; W) \lor (T \; W)} \\
{\Pi}^{{\tau}}_{{(\tau \rightarrow \tte) \rightarrow \tte}} &:= {\lambda P_{\tau\rightarrow\tte}. \lambda W_\worldtype. \forall X_\tau. \; P \; X \; W}
\end{array}\end{equation*}
A HOL term that simulates the original FOML formula is then obtained by recursively
replacing all occurrences of FOML connectives in a FOML formula with their
embedded counterparts.
The truth of $\varphi$ at world $w$ in FOML is reduced to truth of the resulting HOL term
applied to a object of type $\worldtype$ that represents $w$.
Finally, the meta-logical notion of consequence is similarly mapped to HOL predicates \cite{BP13}.

The strength of shallow embeddings not only comes from the reuse of existing classical reasoning
technology, but also from its ability to provide a uniform approach for automating
many different non-classical logics~\cite{Ben19}, here modal logics.
All modal logic systems in the modal logic cube can be embedded by including additional
formulae in the embedding process that restrict the encoded accessibility relation $r^i$
according to the corresponding properties of $R^i$ in the given modal logic.
For example, modal logic \textbf{M} additionally assumes the axiom scheme
$\Box_i \varphi \rightarrow \varphi$ that corresponds to reflexive frames, and the analogous
HOL formula $\forall W_\worldtype. \; \accrel^i \; W \; W$ is included in the embedding process.
For varying domains the definition of universal quantification is adjusted by \emph{relativization},
i.e.\ to include an extra \emph{exists-in-world} predicate $\eiw^\tau$ that mimics the different domains 
\cite{BOR12}. 
Cumulative and decreasing domains are then easily characterized by imposing additional
constraints on $\eiw^\tau$~\cite{GSB17}.

\begin{figure}[tb]
\centering
\caption{Embedding process (picture taken from~\cite{Ste22})
\label{fig:embeddingprocess}}
\includegraphics[width=.95\textwidth,interpolate]{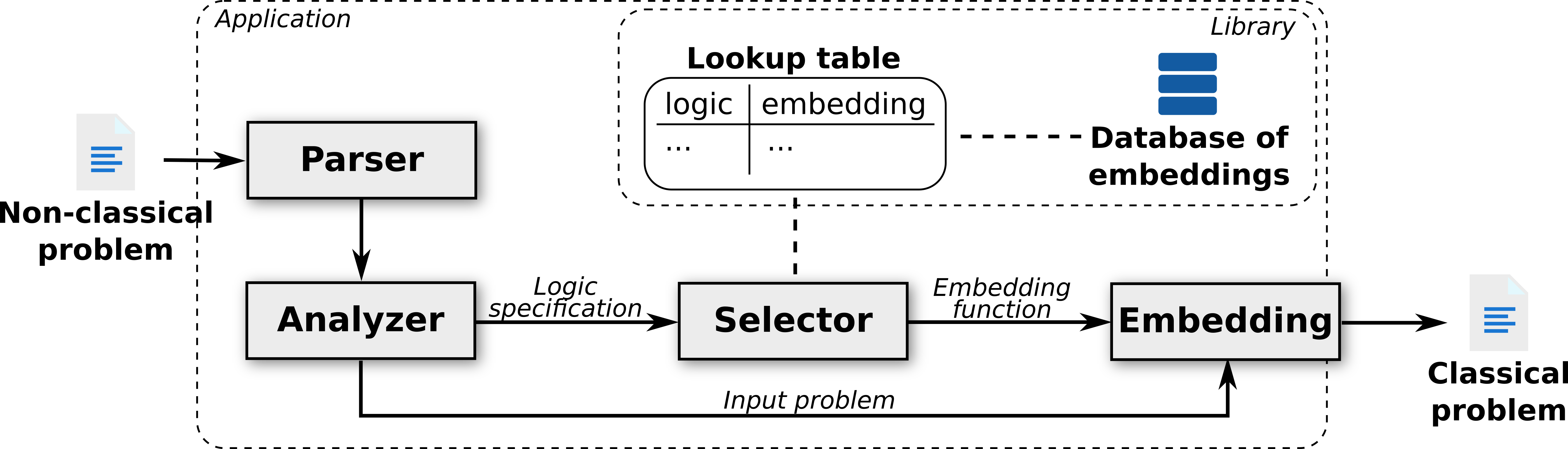}
\end{figure}

The Leo-III system implements the entire embedding process, from reading non-classical logic
problems (including FOML problems) in the NXF and NHF languages, embedding the problems into HOL
using the THF language, and using higher-order reasoning to (attempt to) solve the higher-order
problems.
This is realized using the Logic Embedding Tool \cite{Ste22,Ste24-LE} that implements the embedding
process, as shown in Figure~\ref{fig:embeddingprocess}:
The Logic Embedding Tool will read and parse the input problem formulated in NTF or NHF,
extract its logic specification, choose the right logic embedding function based on the logic specification and
its parameters from a database of available logic embeddings, and finally return the embeddings functions results
in classical TFF or THF. As experimental feature, the tool may also output polymorphically typed result
problems (formulated in TF1 or TH1), though this has not been thoroughly tested and evaluated, yet
(and is not in the scope of this work).
While Leo-III includes the tool as an internal library, the tool is also available as a
stand-alone executable that can be used as external pre-processor with any TPTP-compliant
higher-order ATP system.
It is thus possible to replace Leo-III as the backend reasoning engine, as evaluated in
Section~\ref{Results}.
The embedding tool is open-source and available via Zenodo \cite{Ste24-LE} and GitHub\footnote{%
\href{https://github.com/leoprover/logic-embedding}{\tt https://github.com/leoprover/logic-embedding}}.

\section{Evaluation}
\label{Evaluation}

\subsection{Experimental setup}
\label{Setup}

The QMLTP library v1.1\footnote{%
\href{http://www.iltp.de/qmltp/problems.html}{\tt http://www.iltp.de/qmltp/problems.html}} was
used for the evaluation.
The problems use the normal modal logic box and diamond connectives.
During the preparation of the problems it was noticed that some original QMLTP problems have
syntax errors (missing parentheses, etc.), and some problems use equality without an
axiomatization (including use of the infix {\tt =} predicate that is interpreted in many
ATP systems).
These issues were corrected for the evaluation.\footnote{%
The conditions stated in \cite{RO12} for ``presenting results of modal ATP systems
based on the QMLTP library'' say that ``no part of the problems may be modified''.
As such the results presented in this article for the corrected versions of the QMLTP
problems cannot be called ``results for problems from the QMLTP''.
But pragmatically, the results on the set of (corrected) problems are comparable with 
the results on the original QMLTP problems.}

The sequence of steps taken to prepare the mono-modal problems was:
\begin{itemize}
\item Convert the problems from the QMLTP syntax to the NXF language.
      In particular, the QMLTP {\tt \#box} and {\tt \#dia} connectives were converted to the TPTP's
      {\tt \{\$box\}} and {\tt \{\$dia\}} connectives, and applied to the formula argument.
      The axioms' roles were set to {\tt axiom-local}, to reflect the QMLTP's local consequence.
\item Add each of the 15 logic specifications (three domain types combined with five modal systems) 
      to each of the non-MML NXF problems
      to produce a total of 8700 mono-modal NXF problems.
\item Translate the 8700 NXF problems to both classical TFF and classical THF using the logic embedding tool.
\end{itemize}
In the MML domain, there is just one logic specification for each of the 20 problems because
the multi-modal problems come with a single intended semantics in the QMLTP; the pre-processing steps were
otherwise identical to the mono-modal problems. This produces 20 multi-modal NXF problems,
20 embedded TFF problems, and 20 embedded THF problems.
In total the evaluation set contains 8720 NXF problems for the native modal logic systems,
and their corresponding embeddings embeddings in TFF and THF.

E 3.0.03, Leo-III 1.7.8, Vampire 4.8 were used to prove THF and TFF conjectures, and to disprove 
TFF conjectures -- none of them are able to disprove THF conjectures.
Nitpick 2016 was used to disprove THF conjectures -- that's all it can do.
(An analysis of further higher-order backend systems is presented in earlier work~\cite{SS+23}.)
For the native modal logic ATP systems,
MleanCoP 1.3, which implements modal connection calculi, and
nanoCoP-M 2.0, which implements a non-clausal connection calculus, 
were used to prove and disprove NXF conjectures.
These two systems are the strongest native modal logic ATP systems to 
date~\cite{Ott21,Ott14,BOR12}\footnote{%
During the evaluation it was observed that MleanCoP and nanoCoP-M give conflicting results (i.e., 
one claiming the conjecture to be a theorem, the other claiming it to be a non-theorem) for 11
problems from the GSY and SYM domains.
After manual inspection and cross-validation using results from the embedded variants, this 
suggests an unsoundness of nanoCoP-M. 
As the author's were not successful in contacting the developer of both systems, the need for a 
further investigation of this issue currently remains. 
For the purpose of this evaluation, the 11 conflicting results from nanoCoP-M were removed.
The measured unsoundness of nanoCoP-M, however, renders this system's results an optimistic 
performance estimation, since it's currently unknown how big the impact of the unsoundness
is in the overall evaluation.}
Both are implemented in Prolog; ECLiPSe Prolog version 5.10 is used.
All reasoning systems except Nitpick were run on the StarExec \cite{SST14} Miami cluster
with a 60s wall clock and 480 CPU time limit.
The StarExec Miami computers have an
octa-core Intel Xeon E5-2667 3.20 GHz CPU,
128 GiB memory,
and run the CentOS Linux release 7.4.1708 operating system.
Nitpick was run on a server\footnote{%
Because it could not be installed in StarExec.} with a 60s wall clock time limit.
The server has an
octa-core Intel Xeon E5-2609 2.50 GHz CPU,
64 GiB memory,
and the CentOS Linux release 7.9.2009 operating system.
All source problems used in the evaluation and the evaluation results are available as supplemental 
material to this article via Zenodo~\cite{SSB24}.
The auxiliary scripts for generating the problem files are available from the TPTP's non-classical 
logic GitHub repository\footnote{%
See the QMLTP directory at \url{https://github.com/TPTPWorld/NonClassicalLogic}.}
(recall from Section~\ref{Embedding} that the embedding tool is available separately).

Tables~\ref{DetailedResults1}, ~\ref{DetailedResults2}, and ~\ref{DetailedResultsMML} give the 
numbers of problems solved.
The tables present the numbers of proofs (THM) and disproofs (CSA) for the various logics,
the various logic parameters, the two embeddings (THF and TFF) and their union (T?F), the NXF
problems used by the native modal logic systems, and various unions ($\cup$) of problems solved.
The columns marked with a backslash give the number of problems solved by that union and not
solved by the union of the group indicated after the backslash, e.g., in the T?0 column group,
the \textbackslash NXF column is the number problems solved by the union of the ATP systems on the
THF and TFF embeddings, but not solved by either ATP system on the NXF problems.
The individual ATP systems are abbreviated as ``Leo'' for Leo-III, ``Vam'' for Vampire, ``Nit'' 
for Nitpick, ``MlC'' for ``MleanCoP'', and ``nnC'' for nanoCoP-M.
In each rown the best individual system performance is {\bf bolded}.

\subsection{Results for Mono-Modal Problems}
\label{Results}

Table~\ref{DetailedResults1}, shows the results for modal logics \textbf{M}, \textbf{D},
\textbf{S4} and \textbf{S5}, i.e., the logics supported by the native modal logic systems.
Table~\ref{DetailedResults1} allows for the following observations:
\begin{itemize}
  \item In eight out of the 12 logic configurations MleanCoP proves more theorems than any other 
        system.
        In three configurations Vampire proves the most (using the TFF embedding), and in one 
        configuration E proves the most (using the THF embedding).
  \item MleanCoP generally performs better than nanoCoP-M for proving theorems,
        Vampire is the strongest system for proving theorems in TFF,
        and in THF Leo-III is stronger than the other systems in six configurations.
        Vampire and E are strongest in three configurations each. 
        It is notable that Vampire usually performs better with constant domains.
  \item On average, in \textbf{M} MleanCoP 2.1\% more theorems than the best system using an 
        embedding.
        In \textbf{D} this increases to 3.1\%, and in \textbf{S4} it increases to 6.1\%.
  \item In \textbf{S5} Vampire proves the most theorems (using the TFF embedding), which is 9.3\% 
        more than MleanCoP and 26.0\% more than nanoCoP-M.
        In general, in \textbf{S5}/const and \textbf{S5}/cumul all the embedding-based systems are 
        stronger than the native modal logic systems.
  \item The union of the THF and TFF embedding variants (T?0 in the table) proves only marginally 
        more theorems than either one of them.
  \item Each prover capable of disproving conjectures using either embedding approach is stronger 
        than both the best native modal prover and the union of the native modal logic systems in 
        every logic configuration.
  \item nanoCoP-M generally performs better than MleanCoP for disproving conjectures, in TFF 
        Vampire is always the best system, and in THF Nitpick outperforms every other system in 
        every logic configuration.
  \item On average, in \textbf{M} Nitpick disproves 61.8\% more conjectures than nanoCoP-M.
        In \textbf{D} this drops to 30.2\%, but in \textbf{S4} it increases to 64.2\%, and
        in \textbf{S5} it increases to 78.5\%.
  \item The union of the embedding variants often disproves significantly more conjectures than 
        Nitpick alone.
  \item Both the native modal logic systems and the embedding-based approaches find unique 
        solutions, but only in the embedding approach this is the case for all logic 
        configurations.
  \item Comparing the embedding variants' unions shows that THF yields the strongest results for 
        proving theorems in 10 out of 12 logic configurations.
        For disproving conjectures, the THF variant outperforms the TFF variant in every logic
        configuration.
\end{itemize}

\begin{table}[h!]
\centering
\caption{Mono-modal Results}
\label{DetailedResults1}
\resizebox{\textwidth}{!}{
\begin{tabular}{lll|R{0.75cm}R{0.75cm}R{0.75cm}R{0.75cm}R{0.75cm}R{0.75cm}|R{0.75cm}R{0.75cm}R{0.85cm}R{0.75cm}R{0.75cm}|R{0.75cm}R{0.75cm}|R{0.85cm}R{0.75cm}R{0.75cm}R{0.75cm}|R{0.75cm}|}
\hline
\multicolumn{2}{l}{Language} & SZS & \multicolumn{6}{c|}{THF} & \multicolumn{5}{c|}{TFF} & \multicolumn{2}{c|}{T?0} & \multicolumn{4}{c|}{NXF} & $\cup$ \\
\multicolumn{2}{l}{System}   &     &    E &  Leo &  Vam &   Nit & $\cup$ & {\scriptsize \textbackslash TFF} &    E &  Leo &  Vam &$\cup$ & {\scriptsize \textbackslash THF} & $\cup$ & {\scriptsize \textbackslash NXF} & MlC & nnC & $\cup$ & {\scriptsize \textbackslash T?0} & $\cup$ \\
\hline
\bf M  & const  & THM   &  249 &  251 &  258 &    0 &  262 &    2 &  246 &  251 &  256 &  264 &    4 &  266 &    9 &  {\bf 264} &  236 &  270 &   13 &  279 \\
       &        & CSA   &    0 &    0 &    0 &  {\bf 181} &  181 &   22 &  149 &    0 &  165 &  165 &    6 &  187 &   71 &  110 &  117 &  117 &    1 &  188 \\
       &        &$\cup$ &  249 &  251 &  258 &  181 &  443 &   24 &  395 &  251 &  {\bf 421} &  429 &   10 &  453 &   80 &  374 &  353 &  387 &   14 &  467 \\
\bf M  & vary   & THM   &  199 &  207 &  203 &    0 &  225 &    8 &  193 &  200 &  215 &  221 &    4 &  229 &   12 &  {\bf 217} &  192 &  223 &    6 &  235 \\
       &        & CSA   &    0 &    0 &    0 &  {\bf 243} &  243 &   49 &  178 &    0 &  200 &  200 &    6 &  249 &   93 &  146 &  149 &  156 &    0 &  249 \\
       &        &$\cup$ &  199 &  207 &  203 &  243 &  468 &   57 &  371 &  200 &  {\bf 415} &  421 &   10 &  478 &  105 &  363 &  341 &  379 &    6 &  484 \\
\bf M  & cuml   & THM   &  213 &  230 &  228 &    0 &  248 &    8 &  208 &  220 &  239 &  246 &    6 &  254 &    9 &  {\bf 246} &  217 &  251 &    6 &  260 \\
       &        & CSA   &    0 &    0 &    0 &  {\bf 202} &  202 &   33 &  153 &    0 &  174 &  174 &    5 &  207 &   76 &  124 &  131 &  132 &    1 &  208 \\
       &        &$\cup$ &  213 &  230 &  228 &  202 &  450 &   41 &  361 &  220 &  {\bf 413} &  420 &   11 &  461 &   85 &  370 &  348 &  383 &    7 &  468 \\
\bf M  & $\cup$ & THM   &  661 &  688 &  689 &    0 &  735 &   18 &  647 &  671 &  710 &  731 &   14 &  749 &   30 &  {\bf 727} &  645 &  744 &   25 &  774 \\
       &        & CSA   &    0 &    0 &    0 &  {\bf 626} &  626 &  104 &  480 &    0 &  539 &  539 &   17 &  643 &  240 &  380 &  397 &  405 &    2 &  645 \\
       &        &$\cup$ &  661 &  688 &  689 &  626 & 1361 &  122 & 1127 &  671 & {\bf 1249} & 1270 &   31 & 1392 &  270 & 1107 & 1042 & 1149 &   27 & 1419 \\
\hline
\bf D  & const  & THM   &  195 &  199 &  200 &    0 &  204 &    1 &  193 &  199 &  {\bf 205} &  208 &    5 &  209 &   12 &  203 &  171 &  207 &   10 &  219 \\
       &        & CSA   &    0 &    0 &    0 &  {\bf 302} &  302 &   65 &  199 &    0 &  243 &  243 &    6 &  308 &  102 &  211 &  228 &  228 &   22 &  330 \\
       &        &$\cup$ &  195 &  199 &  200 &  302 &  506 &   66 &  392 &  199 &  {\bf 448} &  451 &   11 &  517 &  114 &  414 &  399 &  435 &   32 &  549 \\
\bf D  & vary   & THM   &  147 &  158 &  148 &    0 &  169 &    8 &  146 &  151 &  158 &  165 &    4 &  173 &    9 &  {\bf 168} &  134 &  170 &    6 &  179 \\
       &        & CSA   &    0 &    0 &    0 &  {\bf 340} &  340 &   90 &  231 &    0 &  253 &  255 &    5 &  345 &   90 &  257 &  265 &  277 &   22 &  367 \\
       &        &$\cup$ &  147 &  158 &  148 &  340 &  509 &   98 &  377 &  151 &  411 &  420 &    9 &  518 &   99 &  {\bf 425} &  399 &  447 &   28 &  546 \\
\bf D  & cuml   & THM   &  162 &  178 &  165 &    0 &  188 &    9 &  162 &  168 &  177 &  185 &    6 &  194 &   10 &  {\bf 187} &  155 &  190 &    6 &  200 \\
       &        & CSA   &    0 &    0 &    0 &  {\bf 319} &  319 &  127 &  170 &    0 &  180 &  197 &    5 &  324 &   96 &  233 &  245 &  251 &   23 &  347 \\
       &        &$\cup$ &  162 &  178 &  165 &  319 &  507 &  136 &  332 &  168 &  357 &  382 &   11 &  518 &  106 &  {\bf 420} &  400 &  441 &   29 &  547 \\
\bf D  & $\cup$ & THM   &  504 &  535 &  513 &    0 &  561 &   18 &  501 &  518 &  540 &  558 &   15 &  576 &   31 &  {\bf 558} &  460 &  567 &   22 &  598 \\
       &        & CSA   &    0 &    0 &    0 &  {\bf 961} &  961 &  282 &  600 &    0 &  676 &  695 &   16 &  977 &  288 &  701 &  738 &  756 &   67 & 1044 \\
       &        &$\cup$ &  504 &  535 &  513 &  961 & 1522 &  300 & 1101 &  518 & 1216 & 1253 &   31 & 1553 &  319 & {\bf 1259} & 1198 & 1323 &   89 & 1642 \\
\hline
\bf S4 & const  & THM   &  290 &  301 &  319 &    0 &  324 &   12 &  288 &  299 &  306 &  316 &    4 &  328 &   11 &  {\bf 331} &  290 &  342 &   25 &  353 \\
       &        & CSA   &    0 &    0 &    0 &  {\bf 133} &  133 &   26 &   91 &    0 &  112 &  112 &    5 &  138 &   55 &   79 &   83 &   83 &    0 &  138 \\
       &        &$\cup$ &  290 &  301 &  319 &  133 &  457 &   38 &  379 &  299 &  {\bf 418} &  428 &    9 &  466 &   66 &  410 &  373 &  425 &   25 &  491 \\
\bf S4 & vary   & THM   &  231 &  247 &  234 &    0 &  264 &    8 &  230 &  239 &  251 &  259 &    3 &  267 &   13 &  {\bf 261} &  225 &  266 &   12 &  279 \\
       &        & CSA   &    0 &    0 &    0 &  {\bf 211} &  211 &   64 &  119 &    0 &  152 &  152 &    5 &  216 &   90 &  118 &  120 &  126 &    0 &  216 \\
       &        &$\cup$ &  231 &  247 &  234 &  211 &  475 &   72 &  349 &  239 &  {\bf 403} &  411 &    8 &  483 &  103 &  379 &  345 &  392 &   12 &  495 \\
\bf S4 & cuml   & THM   &  255 &  278 &  269 &    0 &  297 &    9 &  252 &  269 &  286 &  295 &    7 &  304 &   11 &  {\bf 316} &  278 &  327 &   34 &  338 \\
       &        & CSA   &    0 &    0 &    0 &  {\bf 147} &  147 &   28 &  101 &    0 &  124 &  124 &    5 &  152 &   55 &   92 &   96 &   97 &    0 &  152 \\
       &        &$\cup$ &  255 &  278 &  269 &  147 &  444 &   37 &  353 &  269 &  {\bf 410} &  419 &   12 &  456 &   66 &  408 &  374 &  424 &   34 &  490 \\
\bf S4 & $\cup$ & THM   &  776 &  826 &  822 &    0 &  885 &   29 &  770 &  807 &  843 &  870 &   14 &  899 &   35 &  {\bf 908} &  793 &  935 &   71 &  970 \\
       &        & CSA   &    0 &    0 &    0 &  {\bf 491} &  491 &  118 &  311 &    0 &  388 &  388 &   15 &  506 &  200 &  289 &  299 &  306 &    0 &  506 \\
       &        &$\cup$ &  776 &  826 &  822 &  491 & 1376 &  147 & 1081 &  807 & {\bf 1231} & 1258 &   29 & 1405 &  235 & 1197 & 1092 & 1241 &   71 & 1476 \\
\hline
\bf S5 & const  & THM   &  444 &  434 &  420 &    0 &  455 &    9 &  442 &  433 &  {\bf 446} &  454 &    8 &  463 &   47 &  407 &  355 &  417 &    1 &  464 \\
       &        & CSA   &    0 &    0 &    0 &   {\bf 78} &   78 &    6 &   76 &    0 &   77 &   77 &    5 &   83 &   41 &   38 &   42 &   42 &    0 &   83 \\
       &        &$\cup$ &  444 &  434 &  420 &   78 &  533 &   15 &  518 &  433 &  {\bf 523} &  531 &   13 &  546 &   88 &  445 &  397 &  459 &    1 &  547 \\
\bf S5 & vary   & THM   &  343 &  343 &  326 &    0 &  361 &    6 &  340 &  335 &  {\bf 360} &  366 &   11 &  372 &   34 &  331 &  284 &  339 &    1 &  373 \\
       &        & CSA   &    0 &    0 &    0 &  {\bf 151} &  151 &   22 &  124 &    0 &  133 &  134 &    5 &  156 &   62 &   87 &   88 &   94 &    0 &  156 \\
       &        &$\cup$ &  343 &  343 &  326 &  151 &  512 &   28 &  464 &  335 &  {\bf 493} &  500 &   16 &  528 &   96 &  418 &  372 &  433 &    1 &  529 \\
\bf S5 & cuml   & THM   &  {\bf 449} &  436 &  421 &    0 &  458 &    9 &  442 &  427 &  446 &  456 &    7 &  465 &   48 &  407 &  355 &  417 &    0 &  465 \\
       &        & CSA   &    0 &    0 &    0 &   {\bf 78} &   78 &    6 &   76 &    0 &   77 &   77 &    5 &   83 &   41 &   38 &   42 &   42 &    0 &   83 \\
       &        &$\cup$ &  449 &  436 &  421 &   78 &  536 &   15 &  518 &  427 &  {\bf 523} &  533 &   12 &  548 &   89 &  445 &  397 &  459 &    0 &  548 \\
\bf S5 & $\cup$ & THM   & 1236 & 1213 & 1167 &    0 & 1274 &   24 & 1224 & 1195 & {\bf 1252} & 1276 &   26 & 1300 &  129 & 1145 &  994 & 1173 &    2 & 1302 \\
       &        & CSA   &    0 &    0 &    0 &  {\bf 307} &  307 &   34 &  276 &    0 &  287 &  288 &   15 &  322 &  144 &  163 &  172 &  178 &    0 &  322 \\
       &        &$\cup$ & 1236 & 1213 & 1167 &  307 & 1581 &   58 & 1500 & 1195 & {\bf 1539} & 1564 &   41 & 1622 &  273 & 1308 & 1166 & 1351 &    2 & 1624 \\
\hline
\end{tabular}
}
\end{table}

The presentation of results is continued in Table~\ref{DetailedResults2} for $\textbf{K}$, and 
with summaries over the different logics.
Neither MleanCoP nor nanoCoP-M support reasoning in \textbf{K}.
The results for the THF and TFF systems are included here as their performance can be
approximately compared to the reported results~\cite{BOR12} for MleanSeP 1.2 -- the only
other native ATP system for FOML that supports \textbf{K} and is indexed in the QMLTP.
Previous evaluations indicate that the embedding approach outperforms 
MleanSeP~\cite{BOR12,SS+23}.\footnote{%
The evaluation of MleanSeP in ~\cite{BOR12} uses a higher CPU time limit of 600s, compared to
the 60s used in this work, and the hardware employed in is different. 
Approximate are still possible.}
Since MleanSeP is outperformed in all other logics by MleanCoP and nanoCoP-M~\cite{Ott21},
it is not further considered in the evaluation.

\begin{table}[hbt]
\centering
\caption{Mono-modal Results continued}
\label{DetailedResults2}
\resizebox{\textwidth}{!}{
\begin{tabular}{lll|R{0.75cm}R{0.75cm}R{0.75cm}R{0.85cm}R{0.75cm}R{0.75cm}|R{0.75cm}R{0.75cm}R{0.85cm}R{0.75cm}R{0.75cm}|R{0.75cm}R{0.75cm}|R{0.85cm}R{0.75cm}R{0.75cm}R{0.75cm}|R{0.75cm}|}
\hline
\multicolumn{2}{l|}{Language} & SZS & \multicolumn{6}{c|}{THF} & \multicolumn{5}{c|}{TFF} & \multicolumn{2}{c|}{T?0} & \multicolumn{4}{c|}{NXF} & $\cup$ \\
\multicolumn{2}{l|}{System}   &     &    E &  Leo &  Vam &   Nit & $\cup$ & {\scriptsize \textbackslash TFF} &    E &  Leo &  Vam &$\cup$ & {\scriptsize \textbackslash THF} & $\cup$ & {\scriptsize \textbackslash NXF} & MlC & nnC & $\cup$ & {\scriptsize \textbackslash T?0} & $\cup$ \\
\hline
\bf K  & const  & THM   &  180 &  186 &  184 &    0 &  190 &    1 &  183 &  184 &  {\bf 196} &  202 &   13 &  203 &    - &    - &    - &    - &    - &    - \\
       &        & CSA   &    0 &    0 &    0 &  {\bf 316} &  316 &   38 &  262 &    0 &  296 &  298 &   20 &  336 &    - &    - &    - &    - &    - &    - \\ 
       &        &$\cup$ &  180 &  186 &  184 &  316 &  506 &   39 &  445 &  184 &  {\bf 492} &  500 &   33 &  539 &    - &    - &    - &    - &    - &    - \\
\bf K  & vary   & THM   &  134 &  150 &  140 &    0 &  159 &    8 &  137 &  141 &  {\bf 158} &  165 &   14 &  173 &    - &    - &    - &    - &    - &    - \\
       &        & CSA   &    0 &    0 &    0 &  {\bf 350} &  350 &   86 &  249 &    0 &  267 &  269 &    5 &  355 &    - &    - &    - &    - &    - &    - \\
       &        &$\cup$ &  134 &  150 &  140 &  350 &  509 &   94 &  386 &  141 &  {\bf 425} &  434 &   19 &  528 &    - &    - &    - &    - &    - &    - \\
\bf K  & cuml   & THM   &  150 &  167 &  155 &    0 &  176 &    7 &  151 &  157 &  {\bf 176} &  184 &   15 &  191 &    - &    - &    - &    - &    - &    - \\
       &        & CSA   &    0 &    0 &    0 &  {\bf 330} &  330 &   93 &  220 &    0 &  238 &  242 &    5 &  335 &    - &    - &    - &    - &    - &    - \\
       &        &$\cup$ &  150 &  167 &  155 &  330 &  506 &  100 &  371 &  157 &  {\bf 414} &  426 &   20 &  526 &    - &    - &    - &    - &    - &    - \\
\bf K  & $\cup$ & THM   &  464 &  503 &  479 &    0 &  525 &   16 &  471 &  482 &  {\bf 530} &  551 &   42 &  567 &    - &    - &    - &    - &    - &    - \\
       &        & CSA   &    0 &    0 &    0 &  {\bf 996} &  996 &  217 &  731 &    0 &  801 &  809 &   30 & 1026 &    - &    - &    - &    - &    - &    - \\
       &        &$\cup$ &  464 &  503 &  479 &  996 & 1521 &  233 & 1202 &  482 & {\bf 1331} & 1360 &   72 & 1593 &    - &    - &    - &    - &    - &    - \\
\hline
$\cup$ & const  & THM   & 1358 & 1371 & 1381 &    0 & 1435 &   25 & 1352 & 1366 & {\bf 1409} & 1444 &   34 & 1469 &    - &    - &    - &    - &    - &    - \\
       &        & CSA   &    0 &    0 &    0 & {\bf 1010} & 1010 &  157 &  777 &    0 &  893 &  895 &   42 & 1052 &    - &    - &    - &    - &    - &    - \\
       &        &$\cup$ & 1358 & 1371 & 1381 & 1010 & 2445 &  182 & 2129 & 1366 & {\bf 2302} & 2339 &   76 & 2521 &    - &    - &    - &    - &    - &    - \\
$\cup$ & vary   & THM   & 1054 & 1105 & 1051 &    0 & 1178 &   38 & 1046 & 1066 & {\bf 1142} & 1176 &   36 & 1214 &    - &    - &    - &    - &    - &    - \\
       &        & CSA   &    0 &    0 &    0 & {\bf 1295} & 1295 &  311 &  901 &    0 & 1005 & 1010 &   26 & 1321 &    - &    - &    - &    - &    - &    - \\
       &        &$\cup$ & 1054 & 1105 & 1051 & 1295 & 2473 &  349 & 1947 & 1066 & {\bf 2147} & 2186 &   62 & 2535 &    - &    - &    - &    - &    - &    - \\
$\cup$ & cuml   & THM   & 1229 & 1289 & 1238 &    0 & 1367 &   42 & 1215 & 1241 & {\bf 1324} & 1366 &   41 & 1408 &    - &    - &    - &    - &    - &    - \\
       &        & CSA   &    0 &    0 &    0 & {\bf 1076} & 1076 &  287 &  720 &    0 &  793 &  814 &   25 & 1101 &    - &    - &    - &    - &    - &    - \\
       &        &$\cup$ & 1229 & 1289 & 1238 & 1076 & 2443 &  329 & 1935 & 1241 & {\bf 2117} & 2180 &   66 & 2509 &    - &    - &    - &    - &    - &    - \\
$\cup$ & $\cup$ & THM   & 3641 & 3765 & 3670 &    0 & 3980 &  105 & 3613 & 3673 & {\bf 3875} & 3986 &  111 & 4091 &    - &    - &    - &    - &    - &    - \\
       &        & CSA   &    0 &    0 &    0 & {\bf 3381} & 3381 &  755 & 2398 &    0 & 2691 & 2719 &   93 & 3474 &    - &    - &    - &    - &    - &    - \\
       &        &$\cup$ & 3641 & 3765 & 3670 & 3381 & 7361 &  860 & 6011 & 3673 & {\bf 6566} & 6705 &  204 & 7565 &    - &    - &    - &    - &    - &    - \\
\hline
$\cup${\scriptsize \textbackslash {\bf K}} 
       & const  & THM   & 1178 & 1185 & 1197 &    0 & 1245 &   24 & 1169 & 1182 & {\bf 1213} & 1242 &   21 & 1266 &   79 & 1205 & 1052 & 1236 &   49 & 1315 \\
       &        & CSA   &    0 &    0 &    0 &  {\bf 694} &  694 &  119 &  515 &    0 &  597 &  597 &   22 &  716 &  269 &  438 &  470 &  470 &   23 &  739 \\
       &        &$\cup$ & 1178 & 1185 & 1197 &  694 & 1939 &  143 & 1684 & 1182 & {\bf 1810} & 1839 &   43 & 1982 &  348 & 1643 & 1522 & 1706 &   72 & 2054 \\
$\cup${\scriptsize \textbackslash {\bf K}}
       & vary   & THM   &  920 &  955 &  911 &    0 & 1019 &   30 &  909 &  925 &  {\bf 984} & 1011 &   22 & 1041 &   68 &  977 &  835 &  998 &   25 & 1066 \\
       &        & CSA   &    0 &    0 &    0 &  {\bf 945} &  945 &  225 &  652 &    0 &  738 &  741 &   21 &  966 &  335 &  608 &  622 &  653 &   22 &  988 \\
       &        &$\cup$ &  920 &  955 &  911 &  945 & 1964 &  255 & 1561 &  925 & {\bf 1722} & 1752 &   43 & 2007 &  403 & 1585 & 1457 & 1651 &   47 & 2054 \\
$\cup${\scriptsize \textbackslash {\bf K}}
       & cuml   & THM   & 1079 & 1122 & 1083 &    0 & 1191 &   35 & 1064 & 1084 & 1148 & 1182 &   26 & 1217 &   78 & {\bf 1156} & 1005 & 1185 &   46 & 1263 \\
       &        & CSA   &    0 &    0 &    0 &  {\bf 746} &  746 &  194 &  500 &    0 &  555 &  572 &   20 &  766 &  268 &  487 &  514 &  522 &   24 &  790 \\
       &        &$\cup$ & 1079 & 1122 & 1083 &  746 & 1937 &  229 & 1564 & 1084 & {\bf 1703} & 1754 &   46 & 1983 &  346 & 1643 & 1519 & 1707 &   70 & 2053 \\
$\cup${\scriptsize \textbackslash {\bf K}}
       & $\cup$ & THM   & 3177 & 3262 & 3191 &    0 & 3455 &   89 & 3142 & 3191 & {\bf 3345} & 3435 &   69 & 3524 &  225 & 3338 & 2892 & 3419 &  120 & 3644 \\
       &        & CSA   &    0 &    0 &    0 & {\bf 2385} & 2385 &  538 & 1667 &    0 & 1890 & 1910 &   63 & 2448 &  872 & 1533 & 1606 & 1645 &   69 & 2517 \\
       &        &$\cup$ & 3177 & 3262 & 3191 & 2385 & 5840 &  627 & 4809 & 3191 & {\bf 5235} & 5345 &  132 & 5972 & 1097 & 4871 & 4498 & 5064 &  189 & 6161 \\
\hline
\end{tabular}
}
\end{table}

The main observations from Table~\ref{DetailedResults2} are the following:
\begin{itemize}
  \item In \textbf{K} the TFF embedding proves more theorems than the THF embedding, in every 
        logic configuration. 
        Again, Nitpick outperforms all TFF-based systems for disproving conjectures.
  \item When summarizing over all logics (rows starting with ``$\cup$'' in the table), the THF 
        embedding proves slightly more theorems in varying and cumulative domain configurations, 
        whereas the TFF embedding proves slightly more theorems in constant domain configurations. 
  \item For disproving conjectures the union of the THF-based systems disproves 24.3\% more 
        conjectures than the union of the TFF-based systems.
  \item When summarizing over all logics supported by the native modal logic systems
        (rows starting with ''$\cup${\textbackslash \textbf{K}}'' in the table),
        the union of THF-based systems proves 1.1\% more theorems and disproves 45.0\% more 
        conjectures than the union of native modal logic systems.
        The union of TFF-based systems proves 0.5\% more theorems and disproves 16.1\% more.
  \item The number of problems solved (proved or disproved) by the union of the THF-based systems 
        over all logics except \textbf{K} is 9.3\% larger than the number solved by the union of 
        the TFF-based systems, and 15.3\% larger than the number solved by the union of the native 
        modal logic systems.
\end{itemize}

The evaluation results show that the native modal logic systems prove the most theorems
in two thirds of the individual logics.
Their performance lead is, however, small (at most 6.1\% in modal logics \textbf{S4}).
The embedding approach performs better in \textbf{S5} (9.3\% on average),
and even slightly better when averaged over all logics supported by the native modal
logic systems (10.1\% on average).
For disproving conjectures, the embedding approach is significantly stronger overall. 
This makes the embedding approach the overall stronger approach
wrt.\ the number of problems solved.
The evaluation furthermore shows that automating FOML by embedding into THF is superior to using 
a TFF embedding; only marginally stronger for proving theorems but much stronger for disproving 
conjectures. 
This result may be somewhat counter-intuitive as automated reasoning in TFF is much more developed 
than automated reasoning in THF.

Following the promising results of earlier work~\cite{SS+23}, the embedding process automatically
uses an optimization for embeddings of {\bf S5} logics in the mono-modal context.
Here, the alternative characterization of {\bf S5} employing frames with a universal 
accessibility relation, as opposed to requiring an equivalence relation, is used. 
The two variants are known to be equivalent with respect to provability, and the universal variant 
is more efficient for proof automation.
It can also easily be shown that in mono-modal {\bf S5}, cumulative, constant, and decreasing 
domains coincide. 
Hence, the embedding process can use the more light-weight constant domain encoding of {\bf S5}
problems with cumulative or decreasing domains, thus avoiding complicating the output with 
relativized quantifiers.
The evaluation results substantiate the effectiveness of these optimizations.

In earlier work~\cite{SS+23} the embedding approach could not be used for disproving conjectures 
because the it did not ensure that terms are interpreted locally (which is what the QMLTP assumes). 
This is now properly addressed in the logic embedding tool.
This enables model finding using the logic embedding, but also improved reasoning
performance for proving theorems.

\subsection{Results for Multi-Modal Problems}

The numbers of multi-modal problems solved is shown in Table~\ref{DetailedResultsMML}. 
The table is organized as before.
The results show that the best performing individual system is Vampire, which solves 19 out of 
the 20 problems using the TFF embedding.
MleanCoP and nanoCoP-M solve 17 out of the 19 non-\textbf{K} problems.
Due to the inability of the THF-based systems to disprove conjectures, and Nitpick's
inability to prove conjectures, the best result of any individual THF-based system is 10 
solutions.
However, when considering the union problems solved, the THF-based approach is the only one to 
solve all 20 multi-modal problems. 
Up to the authors' knowledge the THF embedding approach is the first to ever produce
solutions to all the multi-modal problems in the QMLTP.

\begin{table}[hbt]
\centering
\caption{Multi-modal Problems Solved}
\label{DetailedResultsMML}
\resizebox{\textwidth}{!}{
\begin{tabular}{ll|R{0.75cm}R{0.75cm}R{0.75cm}R{0.75cm}R{0.75cm}R{0.75cm}|R{0.75cm}R{0.75cm}R{0.75cm}R{0.75cm}R{0.75cm}|R{0.75cm}R{0.75cm}|R{0.75cm}R{0.75cm}R{0.75cm}R{0.75cm}|R{0.75cm}|}
\hline
\multicolumn{2}{l|}{Language} & \multicolumn{6}{c|}{THF} & \multicolumn{5}{c|}{TFF} & \multicolumn{2}{c|}{T?0} & \multicolumn{4}{c|}{NXF} & $\cup$ \\
\multicolumn{2}{l|}{System}   &    E &  Leo &  Vam & Nit &$\cup$ & {\scriptsize \textbackslash TFF} &    E &  Leo &  Vam &$\cup$ & {\scriptsize \textbackslash THF} & $\cup$ & {\scriptsize \textbackslash NXF} & MlC & nnC & $\cup$ & {\scriptsize \textbackslash T?0} & $\cup$ \\
\hline
\bf K  & THM    &    1 &    1 &    1 &    0 &    1 &    0 &    1 &    1 &    1 &    1 &    0 &    1 &    1 &    - &    - &    - &    - &    - \\
\bf K  & CSA    &    0 &    0 &    0 &    0 &    0 &    0 &    0 &    0 &    0 &    0 &    0 &    0 &    0 &    - &    - &    - &    - &    - \\
\bf K  & $\cup$ &    1 &    1 &    1 &    0 &    1 &    0 &    1 &    1 &    1 &    1 &    0 &    1 &    1 &    - &    - &    - &    - &    - \\
\hline
\bf D  & THM    &    2 &    2 &    2 &    0 &    2 &    0 &    2 &    2 &    2 &    2 &    0 &    2 &    0 &    2 &    2 &    2 &    0 &    2 \\
\bf D  & CSA    &    0 &    0 &    0 &    0 &    0 &    0 &    0 &    0 &    0 &    0 &    0 &    0 &    0 &    0 &    0 &    0 &    0 &    0 \\
\bf D  & $\cup$ &    2 &    2 &    2 &    0 &    2 &    0 &    2 &    2 &    2 &    2 &    0 &    2 &    0 &    2 &    2 &    2 &    0 &    2 \\
\hline
\bf S4 & THM    &    5 &    5 &    5 &    0 &    5 &    0 &    5 &    5 &    5 &    5 &    0 &    5 &    1 &    4 &    4 &    4 &    0 &    5 \\
\bf S4 & CSA    &    0 &    0 &    0 &    7 &    7 &    1 &    5 &    0 &    6 &    6 &    0 &    7 &    0 &    6 &    6 &    6 &    0 &    7 \\
\bf S4 & $\cup$ &    5 &    5 &    5 &    7 &   12 &    1 &   10 &    5 &   11 &   11 &    0 &   12 &    1 &   10 &   10 &   10 &    0 &   12 \\
\hline
\bf S5 & THM    &    2 &    2 &    2 &    0 &    2 &    0 &    2 &    2 &    2 &    2 &    0 &    2 &    0 &    2 &    2 &    2 &    0 &    2 \\
\bf S5 & CSA    &    0 &    0 &    0 &    3 &    3 &    0 &    3 &    0 &    3 &    3 &    0 &    3 &    0 &    3 &    3 &    3 &    0 &    3 \\
\bf S5 & $\cup$ &    2 &    2 &    2 &    3 &    5 &    0 &    5 &    2 &    5 &    5 &    0 &    5 &    0 &    5 &    5 &    5 &    0 &    5 \\
\hline
$\cup$ & THM    &   10 &   10 &   10 &    0 &   10 &    0 &   10 &   10 &   10 &   10 &    0 &   10 &    2 &    8 &    8 &    8 &    0 &   10 \\
$\cup$ & CSA    &    0 &    0 &    0 &   10 &   10 &    1 &    8 &    0 &    9 &    9 &    0 &   10 &    0 &    9 &    9 &    9 &    0 &   10 \\
$\cup$ & $\cup$ &   10 &   10 &   10 &   10 &   20 &    1 &   18 &   10 &   {\bf 19} &   19 &    0 &   20 &    2 &   17 &   17 &   17 &    0 &   20 \\
\hline
$\cup${\scriptsize \textbackslash {\bf K}}
       & THM    &    9 &    9 &    9 &    0 &    9 &    0 &    9 &    9 &    9 &    9 &    0 &    9 &    1 &    8 &    8 &    8 &    0 &    9 \\
$\cup${\scriptsize \textbackslash {\bf K}}
       & CSA    &    0 &    0 &    0 &   10 &   10 &    1 &    8 &    0 &    9 &    9 &    0 &   10 &    0 &    9 &    9 &    9 &    0 &   10 \\
$\cup${\scriptsize \textbackslash {\bf K}}
       & $\cup$ &    9 &    9 &    9 &   10 &   19 &    1 &   17 &    9 &   {\bf 18} &   18 &    0 &   19 &    1 &   17 &   17 &   17 &    0 &   19 \\
\hline
\end{tabular}
}
\end{table}

\section{Conclusion}
\label{Conclusion}

The goals of this work were:
\begin{enumerate}   
\item Test the process of translating first-order modal logic problems into classical first-
      and higher-order logic using a shallow embedding.
\item Evaluate multiple first- and higher-order ATP systems' abilities to solve the embedded
      modal logic problems.
\item Compare the performance of native modal logic ATP systems with the embedding
      approach.
\end{enumerate}
The corresponding findings are:
\begin{enumerate}
\item The embedding process is reliable and successful.
\item For some logics, native modal logic ATP systems prove slightly more theorems than the embedding approach,
      and for some logics the other way around. In sum, however, the embedding approach proves more
      theorems than the native modal logic systems
\item The embedding approach outperforms the native modal logic systems for disproving conjectures.
\item The first-order and higher-order systems prove similar numbers of theorems, and the higher-order
      model finder disproves many more conjectures than the first-order ATP systems.
\item The embedding approach can cope with a wider range of modal logics than the native
      modal systems considered.
\end{enumerate}

\paragraph{Further work.}~\\
As noted in a footnote in Section~\ref{TPTPLanguages}, the development of TPTP
World standards for writing ATP solutions beyond common derivations (e.g., CNF refutations) and 
models (e.g., finite models) is still necessary.
In the context of the embedding approach, one particular issue is converting the TFF and THF proofs
back to suitable proofs in the original non-classical logic. This remains further work.

Building general support for non-classical logics in the TPTP World is current and ongoing work.
The TPTP problem library v9.0.0 will include modal logic problems.
It is expected to be released in the second half of 2024.
Collecting modal logic problems for the TPTP is ongoing work. 
In parallel, the tool chains that support use of these problems will be refined and made easily
accessible.
In particular, the TPTP4X utility \cite{Sut07-CSR} will be extended to output formats for existing
non-classical ATP systems, to provide those systems with a bridge to the TPTP problems until they
adopt the TPTP language natively.
Contemporary systems to bridge to include, e.g., K\raisebox{-3pt}{S}P~\cite{NHD20,PN+21},
nanoCoP-M 2.0 \cite{Ott21}, MleanCoP \cite{Ott14}, MetTeL2 \cite{TSK12}, LoTREC \cite{FF+01},
and MSPASS \cite{HS00-TABLEAUX}.

\bibliographystyle{plain}
\bibliography{Bibliography}

\paragraph{Author information}
\begin{enumerate}
  \item Alexander Steen\\
        University of Greifswald, Germany\\
        \texttt{alexander.steen@uni-greifswald.de} \\
        ORCID: 0000-0001-8781-9462
  \item Geoff Sutcliffe \\
        University of Miami, USA \\
        \texttt{geoff@cs.miami.edu} \\
        ORCID: 0000-0001-9120-3927
  \item Christoph Benzmüller \\
        University of Bamberg and FU Berlin, Germany \\
        \texttt{christoph.benzmueller@uni-bamberg.de} \\
        ORCID: 0000-0002-3392-30935
  \end{enumerate}

\end{document}